\documentclass[12pt]{article}

\usepackage{scicite}

\usepackage{times}

\usepackage[labelfont=bf,labelsep=period]{caption}

\usepackage{amsmath,mathtools}
\usepackage{graphicx,multirow}

\newenvironment{sciabstract}{%
\begin{quote} \bf}
{\end{quote}}

\newcounter{lastnote}
\newenvironment{scilastnote}{%
\setcounter{lastnote}{\value{enumiv}}%
\addtocounter{lastnote}{+1}%
\begin{list}%
{\arabic{lastnote}.}
{\setlength{\leftmargin}{.22in}}
{\setlength{\labelsep}{.5em}}}
{\end{list}}

\title{Discovery of the Interstellar Chiral Molecule Propylene Oxide (CH$_3$CHCH$_2$O)}

\author
{Brett A. McGuire$^{\dagger,1,2,\ast}$ \& P. Brandon Carroll,$^{\dagger,2,\ast}$\\ Ryan A. Loomis,$^{3}$ Ian A. Finneran,$^{2}$ Philip R. Jewell,$^{1}$\\ Anthony J. Remijan,$^{1}$ and Geoffrey A. Blake$^{2,4}$\\
\\
\normalsize{$^{\dagger}$These authors contributed equally to this work.}\\
\normalsize{$^{1}$National Radio Astronomy Observatory, Charlottesville, VA 22903}\\
\normalsize{$^{2}$Division of Chemistry and Chemical Engineering, California Institute of Technology}\\
\normalsize{Pasadena, CA 91125}\\
\normalsize{$^{3}$Department of Astronomy, Harvard University, Cambridge, MA 02138}\\
\normalsize{$^{4}$Division of Geological and Planetary Sciences, California Institute of Technology}\\
\normalsize{Pasadena, CA 91125}\\
\\
\normalsize{$^\ast$To whom correspondence should be addressed; E-mails:  bmcguire@nrao.edu \& pcarroll@caltech.edu.}
}

\date{}

\begin{document} 


\baselineskip24pt


\maketitle


\begin{sciabstract}

Life on Earth relies on chiral molecules, that is, species not superimposable on their mirror images.  This manifests itself in the selection of a single molecular handedness, or homochirality, across the biosphere.  We present the astronomical detection of a chiral molecule, propylene oxide (CH$_3$CHCH$_2$O), in absorption toward the Galactic Center.  Propylene oxide is detected in the gas phase in a cold, extended molecular shell around the embedded, massive protostellar clusters in the Sagittarius B2 star-forming region. This material is representative of the earliest stage of solar system evolution in which a chiral molecule has been found.
\end{sciabstract}
\clearpage



The origin of homochirality is a key mystery in the study of our cosmic origins\cite{Pizzarello:2011fz}. While homochirality is itself evolutionarily advantageous\cite{Lunine:2005tf}, the mechanism for the selection of one iso-energetic enantiomer over another is uncertain. Many routes to homochirality have been proposed through the amplification and subsequent transfer of a small primordial enantiomeric excess (e.e. hereafter).  Disentangling these possible mechanisms requires that we understand the potential sources from which an e.e. may arise.  The oldest material on which e.e. data have been taken in the laboratory are meteoritic samples \cite{Engel:1997ux}, yet, the provenance of this e.e. remains a matter of considerable debate\cite{Glavin:2009dt}. Material in molecular clouds from which planetary systems form is processed through circumstellar disks\cite{Cleeves:2014ty}, and can subsequently be incorporated into planet(esimal)s\cite{Chyba:1990yd}. Thus, a primordial e.e. found in the parent molecular cloud may be inherited by the fledgling system. Constraining the origin of e.e. found in meteorites therefore requires the determination of the possible contributions of primordial e.e., and thus the detection of a chiral molecule in these environments. 

\begin{figure}
\centering
\includegraphics[width=\textwidth]{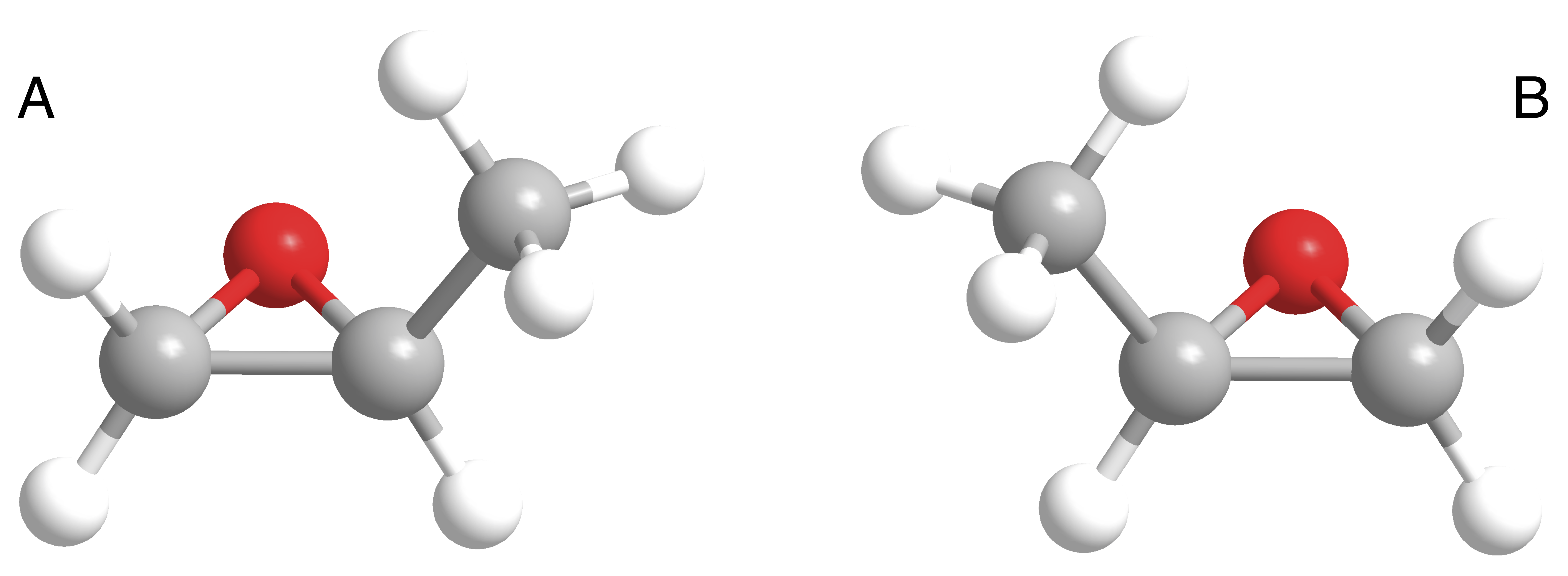}
\captionsetup{labelfont=bf}
\caption{\textbf{The molecular structure of $\textbf{S}$-propylene oxide (A) and $\textbf{R}$-propylene oxide (B)}.  Carbon, hydrogen, and oxygen atoms are indicated by gray, small white, and red spheres, respectively.}
\label{structure}
\end{figure}

For the past fifty years, radio astronomy has been the primary method for studying the gaseous, complex molecular content of interstellar clouds. In this regime, observed spectral features correspond to fine structure transitions of atoms, or pure rotational transitions of polar molecules, that can uniquely identify their carrier. The observations presented here were taken toward the Sagittarius B2 North (Sgr B2(N)) molecular cloud, the pre-eminent source for new complex-molecular detections in the interstellar medium (ISM).

\begin{figure}
\centering
\includegraphics[width=0.5\textwidth]{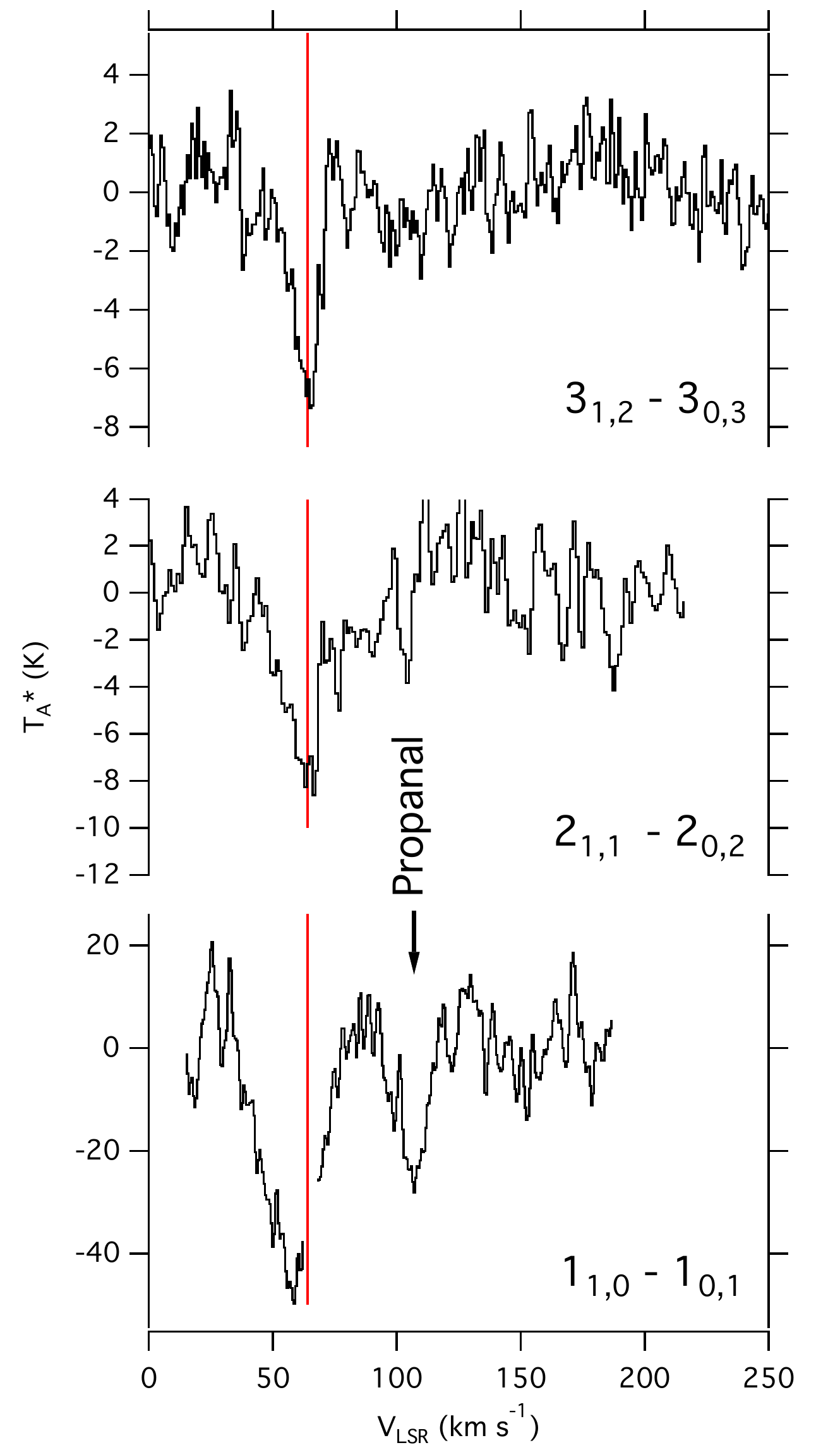}
\captionsetup{labelfont=bf}
\caption{\textbf{Observations of the $\textbf{1}_{1,0} \textbf{- 1}_{0,1}$ (Parkes), $\textbf{2}_{1,1} \textbf{- 2}_{0,2}$ (GBT), and $\textbf{3}_{1,2} \textbf{- 3}_{0,3}$ (GBT) transitions of propylene oxide, in absorption, toward the Galactic Center.}  The 64 km s$^{-1}$ systematic velocity characteristic of Sgr B2(N) is indicated by a vertical red line.  The $1_{0,1} - 1_{1,0}$ transition of propanal is also seen in the Parkes data. Data are given as antenna temperature (T$^*_A$) as a function of shift from local standard of rest velocity (V$_{LSR}$), where 0 km s$^{-1}$ is the measured laboratory frequency of the transition\cite{Science:Materials}, and have been Hanning smoothed.}
\label{observations}
\end{figure}

Propylene oxide (Figure \ref{structure}) was initially detected using data from the publicly-available Prebiotic Interstellar Molecular Survey (PRIMOS) project at the Green Bank Telescope (GBT), which provides near frequency-continuous, high-resolution, high-sensitivity spectral survey data toward Sgr B2(N) from 1 - 50 GHz\cite{Neill:2012fr}.  Based on our model of rotationally-cold propylene oxide absorbing against the Sgr B2(N) continuum\cite{Science:Materials}, only three transitions are predicted to have appreciable intensity above the survey noise floor: the b-type Q-branch $1_{1,0} - 1_{0,1}$, $2_{1,1} - 2_{0,2}$, and $3_{1,2} - 3_{0,3}$ transitions at 12.1, 12.8, and 14.0 GHz ($\lambda = $ 2.478, 2.342, and 2.141 cm), respectively\cite{Science:Materials}.  The $1_{1,0} - 1_{0,1}$ line at 12.1 GHz is obscured by radio frequency interference (RFI) at the GBT, however clear absorption signatures are observed from the $2_{1,1} - 2_{0,2}$, and $3_{1,2} - 3_{0,3}$ transitions (Figure \ref{observations}). 

\begin{figure}
\centering
\includegraphics[width=\textwidth]{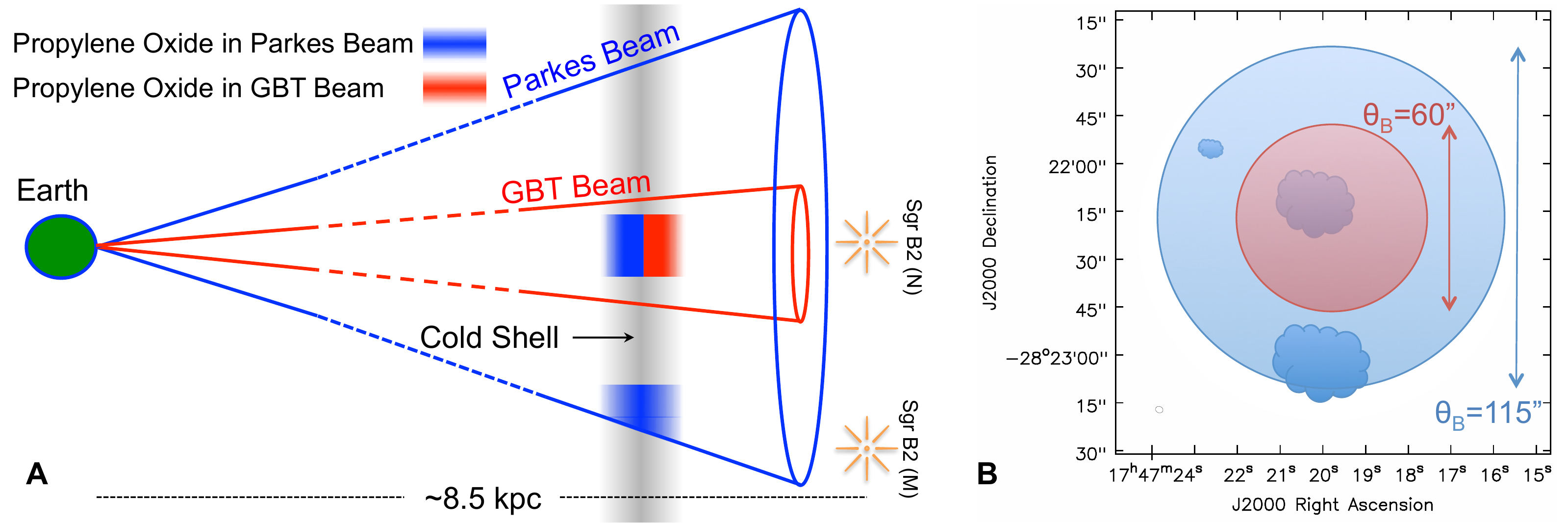}
\captionsetup{labelfont=bf}
\caption{\textbf{Illustration of source structure within the Sgr B2 region.}   A) The GBT and Parkes beam probe different portions of the cold molecular shell in front of the bright continuum sources/hot cores within Sgr B2.  Molecules in the shell which are not backlit by continuum sources are not seen in absorption.  As the schematic of the sky view at right shows, B) The GBT (red) and Parkes (blue) beams probe different continuum sources, with the GBT beam probing only Sgr B2(N), while the Parkes beam also includes most of Sgr B2(M) to the south.}
\label{cartoon}
\end{figure}

These features may be sufficient for a detection on their own at these wavelengths, however we endeavored to confirm the detection by observing the $1_{1,0} - 1_{0,1}$ line at 12.1 GHz using the Parkes Radio Telescope (see Supplementary Materials online text for a detailed description of the Parkes observations), which does not suffer from RFI in the region of the line.  The data confirm the presence of a feature at the same velocity ($\sim$ 64 km s$^{-1}$) as the transitions from PRIMOS, as well as fortuitously detecting a nearby feature of propanal, a structural isomer of propylene oxide (Figure \ref{observations}).  The far-larger Parkes beam ($\sim$115$^{\prime\prime}$ vs 60$^{\prime\prime}$) encompasses a much larger sample of environments, inhomogeneously-broadening the observed transition and incorporating a second, distinct $\sim$46 km s$^{-1}$ component not seen by the GBT beam, but previously observed in the material surrounding Sgr B2 (Figure \ref{cartoon})\cite{Jones:2008bm}.

A fit to the observations using a single excitation temperature for propylene oxide finds a column density of $N_T$ = 1 $\times$ 10$^{13}$ cm$^{-2}$ and a rotational excitation temperature of $T_{ex}$ = 5 K \cite{Science:Materials}.  While an excitation temperature of 5 K is indeed the best-fit to the data, we note that the most rigorous constraint on $T_{ex}$ is from the non-detection at higher-frequencies in PRIMOS, giving an upper limit of $\sim$35 K. Changes in $T_{ex}$ significantly affect $N_T$, and model parameters which fit the data nearly as well are possible for excitation conditions between $T_{ex}$ = 5--35 K.  These models all reproduce the observed features from the GBT and Parkes, and are consistent with the non-detection of propylene oxide at 3 mm; under these conditions, no transitions of propylene oxide would be detectable in the reported observations\cite{Cunningham:2007kl}.  

A search of spectral line catalogues reveals no reasonable interfering transitions from other molecular species.  Propylene oxide is an asymmetric rotor with modest rotational constants, and therefore has numerous ($\sim$450) transitions that fall within the PRIMOS data. For lower excitation temperatures (T$_{ex} \approx$ 10-35 K), at most 80 have have an intensity $\geq$1\% of the strongest predicted line.  Of these, $\sim$13\% are unobservable due to a lack of available receivers at the GBT.  Inspection of the entire PRIMOS data set showed no absorption or emission features attributable to propylene oxide at any of these frequencies but the three listed above, in good agreement with the model and the sensitivity of the survey. 

This detection is complementary to the upper limit placed by \cite{Cunningham:2007kl} on the non-detection of warm, compact propylene oxide at $T_{ex} = 200$ K toward Sgr B2(N) at mm-wavelengths using the Mopra Telescope.  This search was sensitive only to a warm population of propylene oxide, however, and resulted in a non-detection with an upper limit column density of $6.7 \times 10^{14}$ cm$^{-2}$ for an excitation temperature of $T_{ex} = 200$ K and compact source size (5$''$) such as that expected for gas associated with the embedded protostellar clusters/hot cores in this cloud\cite{Cunningham:2007kl}.  

In sources with strong background continuum, of which Sgr B2(N) is a prominent example, many rotationally-cold, high dipole moment species are observed almost exclusively in absorption against the continuum source, as shown in Figure \ref{cartoon}. Because of the exceptionally low line densities, only two to five well-measured centimeter-wavelength spectral features are needed to securely claim a detection (see, e.g., \cite{Hollis:2004uh,McGuire:2012jf,Hollis:2004oc}). This stands in stark contrast to mm-wave detections, particularly toward Sgr B2(N), where dozens of lines are typically required. Based on a statistical analysis of the line density in our observations of Sgr B2(N), we find that the likelihood of three random features falling within three resolution elements of the propylene oxide transitions to be $\leq$ 6$\times$10$^{-8}$\cite{Science:Materials}.


Taken together, the GBT and Parkes observations provide strong evidence of cold, low-abundance propylene oxide toward the Sgr B2 cloud complex, in excellent agreement with previously established upper limits, as well as with previous observations of complex organic molecules. Indeed, many of the complex organics seen toward Sgr B2(N) are found not in or near the hot cores, but, like propylene oxide, in a cold, extended shell around the source.  In these regions, molecules are often liberated into the gas phase via non-thermal, shock-driven, desorption, resulting in colder, spatially-extended gas-phase populations that are often more abundant than predicted by standard warm-up models \cite{RequenaTorres:2006ki}.  This is consistent with the observation that the structurally-similar ethylene oxide is consistently found to have low excitation temperatures (11 -- 35 K), well below the temperature of the surrounding grains\cite{Ikeda:2001vd}, with the detections of glycolaldehyde\cite{Hollis:2004uh}, ethanimine\cite{Loomis:2013fs}, and propylene oxide's structural isomers propanal\cite{Hollis:2004oc} and acetone\cite{Science:Materials} in this region, and with the general pattern of shock-driven liberation of complex molecules in the so-called central molecular zone\cite{RequenaTorres:2006ki}.

From a chemical perspective, the presence of propylene oxide in Sgr B2(N) is not surprising. Propylene oxide is the third species of the C$_3$H$_6$O family detected toward this source. Its structural isomers, propanal  (CH$_3$CH$_2$CHO) \cite{Hollis:2004oc} and acetone ((CH$_3$)$_2$CO) \cite{Snyder:2002jd}, are both seen toward Sgr B2(N), and propylene oxide is not the first epoxide found in the ISM.  Ethylene oxide (CH$_2$OCH$_2$) is structurally similar to propylene oxide, differing by only a methyl group, and has been detected toward numerous massive star-forming regions including Sagittarius B2(N)\cite{Dickens:1997fb,Ikeda:2001vd}. In the case of acetone, \cite{Belloche:2013eb} report a column density of $N_T = 1.49 \times 10^{17}$ cm$^{-2}$, but for a warm population with $T_{ex} = 100$ K that peaks at the position of the hot core.  In the detection of propanal, column densities were not determined\cite{Hollis:2004oc}. 

To determine the relative populations of these molecules in the cold shell around Sgr B2(N), we have used the same procedure as for propylene oxide \cite{Science:Materials}. We find that a column density of $N_T = 6 \times 10^{13}$ cm$^{-2}$ with $T_{ex} = 6.2$ K reproduces the 11 propanal  transitions observed in the full PRIMOS dataset (Supplementary online text), to within a factor of $\sim$2. Similarly, using 18 detected lines of acetone in PRIMOS (see Supplementary Materials), we find a column density of $N_T = 2.1 \times 10^{14}$ cm$^{-2}$ with $T_{ex} = 6.2$ K reproduces these features within a factor of $\sim$2.  The best-fit $T_{ex} = 5$ K for propylene oxide is in remarkably good agreement with these values, which due to the larger number of observed transitions over a wider frequency range for propanal and acetone, are much more rigorously-constrained.  $T_{ex}$ up to 35 K for propylene oxide are formally allowed in the propylene oxide fit, due to loose constraints stemming from the narrow range of energy levels covered in a narrow frequency window.  However, the best-fit $T_{ex}$ = 5 K is significantly bolstered by the similar conditions exhibited by the acetone and propanal populations.

All three members of the C$_3$H$_6$O family are then detected in absorption in the PRIMOS data at remarkably similar excitation conditions, suggesting they likely occupy the same cold, shocked region surrounding Sgr B2(N). Propanal and acetone are thermodynamically favored over propylene oxide, residing 22.7 and 30.8 kcal mol$^{-1}$ (0.98 and 1.33 eV) lower in energy, respectively \cite{Lifshitz:1994ik}. But, while the relative column densities derived here do roughly follow the pattern of increasing abundance with increasing stability, chemistry in molecular clouds is largely kinetically-controlled, rather than thermodynamically, and relative abundances do not regularly follow thermodynamic patterns \cite{Herbst:2009go,Loomis:2015jh}. The recent detections of acetone and propanal at an abundance ratio of three to one in comet 67P/Churyumov-Gerasimenko show that members of the C$_3$H$_6$O family also feature prominently in the volatile organic content of comet nuclei, and the remarkably similar ratios to those observed toward Sgr B2(N) suggest that such kinetically-controlled routes to both species are widespread, and not isolated to extraordinary interstellar sources\cite{Goesmann:2015gv}.

The leading models for the production and enhancement of an e.e. in the interstellar medium likely act over timescales far longer than the delivery of complex organic material to the planet-forming region of disks\cite{Modica:2014en,Dreiling:2014ez,Bailey:1998us}.   A number of mechanisms have been proposed for gas-phase routes in the ISM to create such a primordial e.e.  While beta decay-related chemistry has been proven to generate slight chiral asymmetries \cite{Dreiling:2014ez} that would be universal in nature, perhaps the most intriguing route, astronomically, is enantiomerically-selective photochemistry induced by circularly-polarized light (CPL) \cite{Modica:2014en}. Here, the chirally-sensitive chemical reaction networks would be stochastically driven on the spatial scales of giant molecular cloud complexes. Toward the Orion Nebula cluster, for example, significant CPL patterns capable of producing e.e. do not extend over the entire protostellar cluster, but have been detected over regions large compared to individual protoplanetary disks \cite{Bailey:1998us}.  We have rigorously examined the possible mechanisms for determining an e.e. (Supplemental Materials), and concluded that the standard, total power observations shown here cannot determine whether such an e.e. exists in the case of propylene oxide, but that high precision, full polarization state measurements can, in principle.  Critically, the detection of gas-phase propylene oxide toward the Galactic Center provides a molecular target for such observations, and demonstrates that interstellar chemistry can reach sufficient levels of complexity to form chiral species in environments with the physical conditions required to produce an enantiomeric excess.


\bibliography{science}

\begin{thebibliography}{10}

\bibitem{Pizzarello:2011fz}
S.~Pizzarello, T.~L. Groy, {\it Geochimica et Cosmochimica Acta\/} {\bf 75},
  645 (2011).

\bibitem{Lunine:2005tf}
J.~Lunine, {\it {Astrobiology: A Multi-Disciplinary Approach}\/} (Pearson
  Education Inc., San Francisco, CA, 2005).

\bibitem{Engel:1997ux}
M.~H. Engel, S.~A. Macko, {\it Nature\/} {\bf 389}, 265 (1997).

\bibitem{Glavin:2009dt}
D.~P. Glavin, J.~P. Dworkin, {\it Proceedings of the National Academy of
  Sciences of the United States of America\/} {\bf 106}, 5487 (2009).

\bibitem{Cleeves:2014ty}
L.~I. Cleeves, {\it et~al.\/}, {\it Science\/} {\bf 345}, 1590 (2014).

\bibitem{Chyba:1990yd}
C.~F. Chyba, P.~J. Thomas, L.~Brookshaw, C.~Sagan, {\it Science\/} {\bf 249},
  366 (1990).

\bibitem{Neill:2012fr}
J.~L. Neill, {\it et~al.\/}, {\it The Astrophysical Journal\/} {\bf 755}, 153
  (2012).

\bibitem{Science:Materials}
{Materials and methods are available as supplementary materials on
  \emph{Science} Online}.

\bibitem{Jones:2008bm}
P.~A. Jones, {\it et~al.\/}, {\it Monthly Notices of the Royal Astronomical
  Society\/} {\bf 386}, 117 (2008).

\bibitem{Cunningham:2007kl}
M.~R. Cunningham, {\it et~al.\/}, {\it Monthly Notices of the Royal
  Astronomical Society\/} {\bf 376}, 1201 (2007).

\bibitem{Hollis:2004uh}
J.~M. Hollis, P.~R. Jewell, F.~J. Lovas, A.~Remijan, {\it The Astrophysical
  Journal\/} {\bf 613}, L45 (2004).

\bibitem{McGuire:2012jf}
B.~A. McGuire, {\it et~al.\/}, {\it The Astrophysical Journal\/} {\bf 758}, L33
  (2012).

\bibitem{Hollis:2004oc}
J.~M. Hollis, P.~R. Jewell, F.~J. Lovas, A.~Remijan, H.~M{\o}llendal, {\it The
  Astrophysical Journal\/} {\bf 610}, L21 (2004).

\bibitem{RequenaTorres:2006ki}
M.~A. Requena-Torres, {\it et~al.\/}, {\it A{\&}A\/} {\bf 455}, 971 (2006).

\bibitem{Ikeda:2001vd}
M.~{Ikeda}, {\it et~al.\/}, {\it The Astrophysical Journal\/} {\bf 560}, 792
  (2001).

\bibitem{Loomis:2013fs}
R.~A. Loomis, {\it et~al.\/}, {\it The Astrophysical Journal\/} {\bf 765}, L9
  (2013).

\bibitem{Snyder:2002jd}
L.~E. Snyder, {\it et~al.\/}, {\it The Astrophysical Journal\/} {\bf 578}, 245
  (2002).

\bibitem{Dickens:1997fb}
J.~E. Dickens, {\it et~al.\/}, {\it The Astrophysical Journal\/} {\bf 489}, 753
  (1997).

\bibitem{Belloche:2013eb}
A.~{Belloche}, H.~S.~P. {M{\"u}ller}, K.~M. {Menten}, P.~{Schilke},
  C.~{Comito}, {\it Astronomy and Astrophysics\/} {\bf 559}, A47 (2013).

\bibitem{Lifshitz:1994ik}
A.~Lifshitz, C.~Tamburu, {\it The Journal of Physical Chemistry\/} {\bf 98},
  1161 (1994).

\bibitem{Herbst:2009go}
E.~Herbst, E.~F. van Dishoeck, {\it Annual Reviews of Astronomy and
  Astrophysics\/} {\bf 47}, 427 (2009).

\bibitem{Loomis:2015jh}
R.~A. Loomis, {\it et~al.\/}, {\it The Astrophysical Journal\/} {\bf 799}, 34
  (2015).

\bibitem{Goesmann:2015gv}
F.~{Goesmann}, {\it et~al.\/}, {\it Science\/} {\bf 349} (2015).

\bibitem{Modica:2014en}
P.~Modica, {\it et~al.\/}, {\it The Astrophysical Journal\/} {\bf 788}, 79
  (2014).

\bibitem{Dreiling:2014ez}
J.~M. Dreiling, T.~J. Gay, {\it Physical Review Letters\/} {\bf 113}, 118103
  (2014).

\bibitem{Bailey:1998us}
J.~Bailey, {\it et~al.\/}, {\it Science\/} {\bf 281}, 672 (1998).

\bibitem{Swalen:1957fn}
J.~D. Swalen, D.~R. Herschbach, {\it The Journal of Chemical Physics\/} {\bf
  27}, 100 (1957).

\bibitem{Herschbach:1958fa}
D.~R. Herschbach, J.~D. Swalen, {\it The Journal of Chemical Physics\/} {\bf
  29}, 761 (1958).

\bibitem{Finneran:2013td}
I.~A. Finneran, D.~B. Holland, P.~B. Carroll, G.~A. Blake, {\it Review of
  Scientific Instruments\/} {\bf 84}, 3104 (2013).

\bibitem{Zaleski:2013bc}
D.~P. Zaleski, {\it et~al.\/}, {\it The Astrophysical Journal\/} {\bf 765}, L10
  (2013).

\bibitem{Gordy:1984uy}
W.~Gordy, R.~L. Cook, {\it {Microwave Molecular Spectra}\/} (Wiley, New York,
  1984), third edn.

\bibitem{Hollis:2006ih}
J.~M. Hollis, A.~J. Remijan, P.~R. Jewell, F.~J. Lovas, {\it The Astrophysical
  Journal\/} {\bf 642}, 933 (2006).

\bibitem{Townes:1975ve}
C.~H. Townes, {\it {Microwave Spectroscopy}\/} (Dover Publications, New York,
  1975).

\bibitem{Hollis:2007ww}
J.~M. Hollis, P.~R. Jewell, A.~J. Remijan, F.~J. Lovas, {\it The Astrophysical
  Journal\/} {\bf 660}, L125 (2007).

\bibitem{Ladd:2013hn}
N.~Ladd, C.~Purcell, T.~Wong, S.~Robertson, {\it Publications of the
  Astronomical Society of Australia\/} {\bf 22}, 62 (2013).

\bibitem{Mehringer:1993dd}
D.~M. Mehringer, P.~Palmer, W.~M. Goss, F.~Yusef-Zadeh, {\it Astrophysical
  Journal\/} {\bf 412}, 684 (1993).

\bibitem{Crockett:2014er}
N.~R. Crockett, {\it et~al.\/}, {\it The Astrophysical Journal\/} {\bf 787},
  112 (2014).

\bibitem{Snyder:2005tr}
L.~Snyder, {\it et~al.\/}, {\it Astrophys. J.\/} {\bf 619}, 914 (2005).

\bibitem{Salzman:1998jd}
W.~Salzman, {\it Journal of Molecular Spectroscopy\/} {\bf 192}, 61 (1998).

\bibitem{Mason:2009dr}
B.~S. Mason, T.~Robishaw, C.~Heiles, D.~Finkbeiner, C.~Dickinson, {\it The
  Astrophysical Journal\/} {\bf 697}, 1187 (2009).

\bibitem{Robishaw:2009hu}
T.~Robishaw, C.~Heiles, {\it Publications of the Astronomical Society of the
  Pacific\/} {\bf 121}, 272 (2009).

\bibitem{2011piim.book.....D}
B.~T. {Draine}, {\it {Physics of the Interstellar and Intergalactic Medium}\/}
  (2011).

\bibitem{Goldreich:1981ab}
P.~{Goldreich}, N.~D. {Kylafis}, {\it The Astrophysical Journal Letters\/} {\bf
  243}, L75 (1981).

\bibitem{2008ApJ...680..981R}
T.~{Robishaw}, E.~{Quataert}, C.~{Heiles}, {\it The Astrophysical Journal\/}
  {\bf 680}, 981 (2008).

\bibitem{Wagniere:1999ab}
G.~H. {Wagni{\`e}re}, {\it Chemical Physics\/} {\bf 245}, 165 (1999).

\bibitem{Crutcher:1996ab}
R.~M. {Crutcher}, D.~A. {Roberts}, D.~M. {Mehringer}, T.~H. {Troland}, {\it The
  Astrophysical Journal Letters\/} {\bf 462}, L79 (1996).

\bibitem{Bass:2009:HOT:1594759}
M.~Bass, {\it et~al.\/}, {\it Handbook of Optics, Third Edition Volume I:
  Geometrical and Physical Optics, Polarized Light, Components and
  Instruments(Set)\/} (McGraw-Hill, Inc., New York, NY, USA, 2010), third edn.

\end{thebibliography}

\bibliographystyle{Science}


\begin{scilastnote}
\item[] \textbf{Supplementary Materials}

\vspace{-1em}
www.sciencemag.org

\vspace{-1em}
Materials and Methods

\vspace{-1em}
Figures S1, S2, S3, S4, S5

\vspace{-1em}
Tables S1, S2, S3, S4

\vspace{-1em}
References (27-47)
\item[] \textbf{Acknowledgements}
\item[] We thank the ongoing support of the PRIMOS team, GBT, and Parkes staff in acquiring the GBT and Parkes data, and S. Breen, S. Mader, and J. Reynolds for assistance with Parkes data reduction.  We acknowledge the support of L. Snyder, J.M. Hollis, and F. Lovas. B.A.M. thanks J. Mangum and J. Corby for helpful discussions.  P.B.C. and B.A.M. acknowledge the support of an NASA Astrobiology Institute Early Career Collaboration Award. B.A.M. is funded by a National Radio Astronomy Observatory Jansky Postdoctoral Fellowship. R.A.L. and I.A.F. are funded by a National Science Foundation Graduate Research Fellowship. P.B.C., I.A.F. And  G.A.B. acknowledge support from the NASA Astrobiology Institute through the Goddard Team (M. J. Mumma, PI) under Cooperative Research Agreements NNX09AH63A and NNX15AT33A(NNX09AH63A), and the NSF Astronomy and Astrophysics (AST-1109857) grant program. 
Access to the entire PRIMOS data set, specifics on the observing strategy, and overall frequency coverage information is available at http://www.cv.nrao.edu/PRIMOS/.  The spectra obtained with Parkes are available through this website as well.  Data from project AGBT06B-006 are available in the NRAO Archive at https://science.nrao.edu/observing/data-archive.

The National Radio Astronomy Observatory is a facility of the National Science Foundation operated under cooperative agreement by Associated Universities, Inc.

The Australia Telescope Compact Array (/ Parkes radio telescope / Mopra radio telescope / Long Baseline Array) is part of the Australia Telescope National Facility which is funded by the Australian Government for operation as a National Facility managed by CSIRO.
\clearpage
\clearpage

Supplementary Materials for

\title{Discovery of the Interstellar Chiral Molecule Propylene Oxide (CH$_3$CHCH$_2$O)}


\author
{Brett A. McGuire$^{\dagger,1,2,\ast}$ \& P. Brandon Carroll,$^{\dagger,2,\ast}$\\ Ryan A. Loomis,$^{3}$ Ian A. Finneran,$^{2}$ Philip R. Jewell,$^{1}$\\ Anthony J. Remijan,$^{1}$ and Geoffrey A. Blake$^{2,4}$\\
\\
\normalsize{$^{\dagger}$These authors contributed equally to this work.}\\
\normalsize{$^{1}$National Radio Astronomy Observatory, Charlottesville, VA 22903}\\
\normalsize{$^{2}$Division of Chemistry and Chemical Engineering, California Institute of Technology}\\
\normalsize{Pasadena, CA 91125}\\
\normalsize{$^{3}$Department of Astronomy, Harvard University, Cambridge, MA 02138}\\
\normalsize{$^{4}$Division of Geological and Planetary Sciences, California Institute of Technology}\\
\normalsize{Pasadena, CA 91125}\\
\\
\normalsize{$^\ast$To whom correspondence should be addressed; E-mails:  bmcguire@nrao.edu \& pcarroll@caltech.edu.}
}

\item[] \textbf{This PDF file includes:}


\vspace{-1em}
Materials and Methods

\vspace{-1em}
Figures S1, S2, S3, S4, S5

\vspace{-1em}
Tables S1, S2, S3, S4

\vspace{-1em}
References (27-47)

\end{scilastnote}

\clearpage

\part*{Materials \& Methods}

\renewcommand{\thefigure}{S\arabic{figure}}
\renewcommand{\thetable}{S\arabic{table}}
\setcounter{figure}{0}

\section*{Observations}

The full observational details, data reduction strategy, and analysis of the PRIMOS observations presented here have been previously reported (\emph{13}), and will not be discussed further. PRIMOS provides near-continuous frequency coverage of Sgr B2(N) from 1 -- 50 GHz at high sensitivity (RMS$\sim$3 -- 9 mK) and spectral resolution ($\Delta\nu\sim$ 25 kHz) using the 100-m Robert C. Byrd Green Bank Telescope (GBT).  PRIMOS data are fully-reduced and made publicly available with no propriety period.  More information on how to obtain full data sets is available online at http://www.cv.nrao.edu/$\sim$aremijan/PRIMOS/, and large portions of the survey, as well as many others, are available through the \textbf{S}pectral \textbf{Li}ne \textbf{S}earch \textbf{E}ngine (SLiSE) at http://www.cv.nrao.edu/$\sim$aremijan/SLiSE/.  Additionally, several hours of archival GBT observations (GBT Project AGBT06B-006) fortuitously covered the 12.8 GHz transition of propylene oxide.  These data were reduced in the same way as the PRIMOS data, and used to further bolster the signal-to-noise of the line.

Observations with the Parkes Radio Telescope were conducted over a total of 9 nights from 30 April 2015 -- 19 May 2015.  The target coordinates for the observations were the same as those for the PRIMOS observations: right ascension = 17$^{h}$47$^{m}$19.8$^{s}$, declination = -28$^{\circ}$22$^{\prime}$17.0$^{\prime\prime}$ (J2000).  Spectra were acquired in position-switching mode, with the off-position located 1$^{\circ}$ offset in latitude with a switching cycle of 5 minutes.  Pointing accuracy was checked by facility staff and converged to $\sim$5$^{\prime\prime}$ accuracy.  The Parkes $K_u$-band receiver was used with a system temperature of $\sim$80 K across the band.  The backend was an 8 MHz bandwidth, 1 kHz resolution Digital Filter Bank.  

The quasar ICRF J193925.0-634245 was used for absolute flux calibration.  Scans were collected in dual-polarization mode; these were averaged to increase the sensitivity of the observations.  Background flux was removed from the scans using a 5th-order polynomial fit to the baseline.  The resulting spectra were then binned and Hanning smoothed to a resolution of $\sim$0.6 km s$^{-1}$.  Finally, an instrumental noise feature, or `birdy,' was removed at $\sim$12069.8 MHz.

For such weak features there is a possibility that even small systematic response (e.g. standing waves) may obscure or even be mistaken for genuine molecular absorption.  In order to ensure all absorption features were real, each feature was re-observed at a shifted rest frequency such that the target frequency was shifted within the passband. For Parkes observations, these tests were carried out concurrently with the observations described above in the same manner. The additional GBT observations were carried out over four sessions from 20 August 2015 -- 16 September 2015 (GBT project AGBT15A-493). Observations were conducted in position-switching mode, with the off-position located 1$^{\circ}$ offset in azimuth with a switching cycle of 4 minutes.  Pointing accuracy was checked and corrected every $\sim$ 2 hrs using the nearby continuum source PKSJ 1833-2103.  The $2_{11} - 2_{02}$ and $3_{12} - 3_{03}$ lines were observed simultaneously using the GBT $K_u$-band receiver and the VEGAS spectrometer in its 187.5 MHz bandwidth mode. For both Parkes and GBT observations, the targeted transition(s) were successfully reproduced, and were appropriately shifted within the passband. Therefore, the features cannot be attributed to IF or passband response, and must be from the target source at the specified rest frequencies.

\section*{Laboratory Spectroscopy}

The cm-wave spectrum of propylene oxide is complicated by coupling of the internal rotation of the methyl group to the overall rotation of the molecule. This splits each rotational level into a non-degenerate A state and a doubly degenerate E state. The high barrier to internal rotation (2710 cal/mole; 0.1175 eV) produces small splittings, $\sim$ 100 kHz, for low J transitions such as those observed here. These internal rotation-driven splittings will be unresolved for astronomical observations toward Sgr B2(N), but will contribute appreciably to the observed line width. The pure rotational spectrum of propylene oxide has been previously measured and assigned\cite{Swalen:1957fn,Herschbach:1958fa}. However, due to linewidth and resolution limits in this (previous) work, the treatment of the splittings was insufficient for determining the contribution of this splitting to the astronomical linewidth.  We have therefore re-measured these spectra using a state-of-the-art free jet microwave spectrometer.

Specifically, laboratory spectra were collected from 8--18 GHz using a direct digital synthesis (DDS)-based, chirped-pulse Fourier transform microwave (CP-FTMW) spectrometer. Details of the instrument have been published elsewhere\cite{Finneran:2013td}. Briefly, a linear frequency sweep (0~--~2 GHz,~1~$\mu$s duration) chirped pulse is generated by a DDS card. This pulse is then upconverted by mixing with a local oscillator (8--18 GHz), amplified (to 50 W), and broadcast by a waveguide horn into the chamber to interact with the sample. Propylene oxide, sometimes referred to as methyl oxirane, ($>$99\% purity), was purchased from Alfa Aesar, and used without further purification. The sample was prepared by flowing 1 atm of Ar through a sealed reservoir containing $\sim$5 mL of propylene oxide, and introduced into the vacuum chamber through a pulsed adiabatic expansion. 

After excitation by the microwave pulse, the weak molecular free induction decay (FID) was collected using the same microwave horn, amplified (+38 dB), downconverted, and recorded with a high speed digital-to-analog converter (4 GSa/s). The expansion coupled with the coaxial excitation-detection geometry employed produces a rotationally cold ($\sim$3 K) gas with a slight asymmetry in the blended Doppler doublet. This limits the frequency uncertainty in the measured line centers to 20 kHz.

\begin{figure}[h!]
\centering
\includegraphics[width=\textwidth]{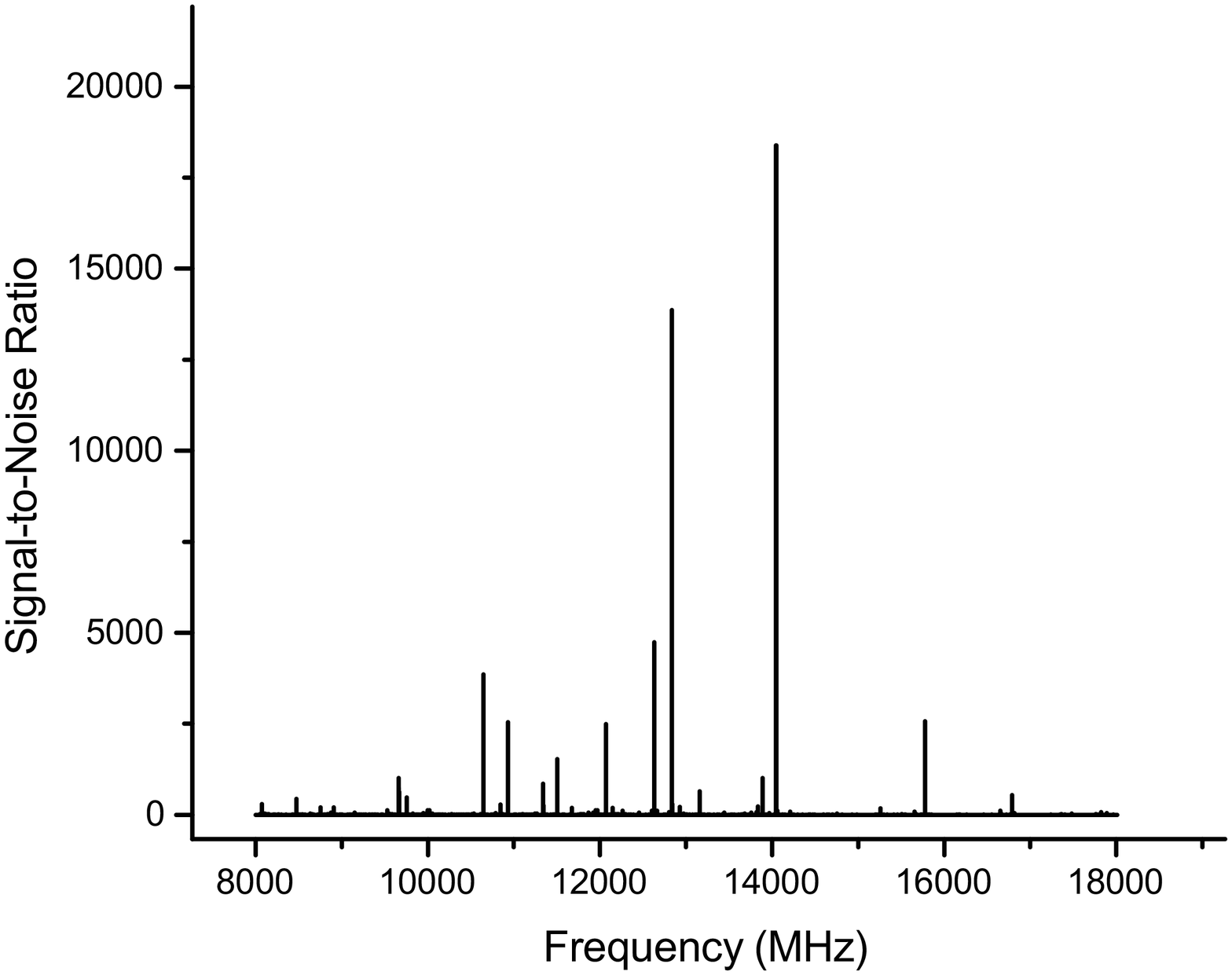}
\caption{\textbf{Laboratory spectrum of propylene oxide.} The spectrum of propylene oxide in an Ar buffer gas from 8 -- 18 GHz (4.8 million averages, 20 hour acquisition)}
\label{propox}
\end{figure}

\begin{figure}
\centering
\includegraphics[width=\textwidth]{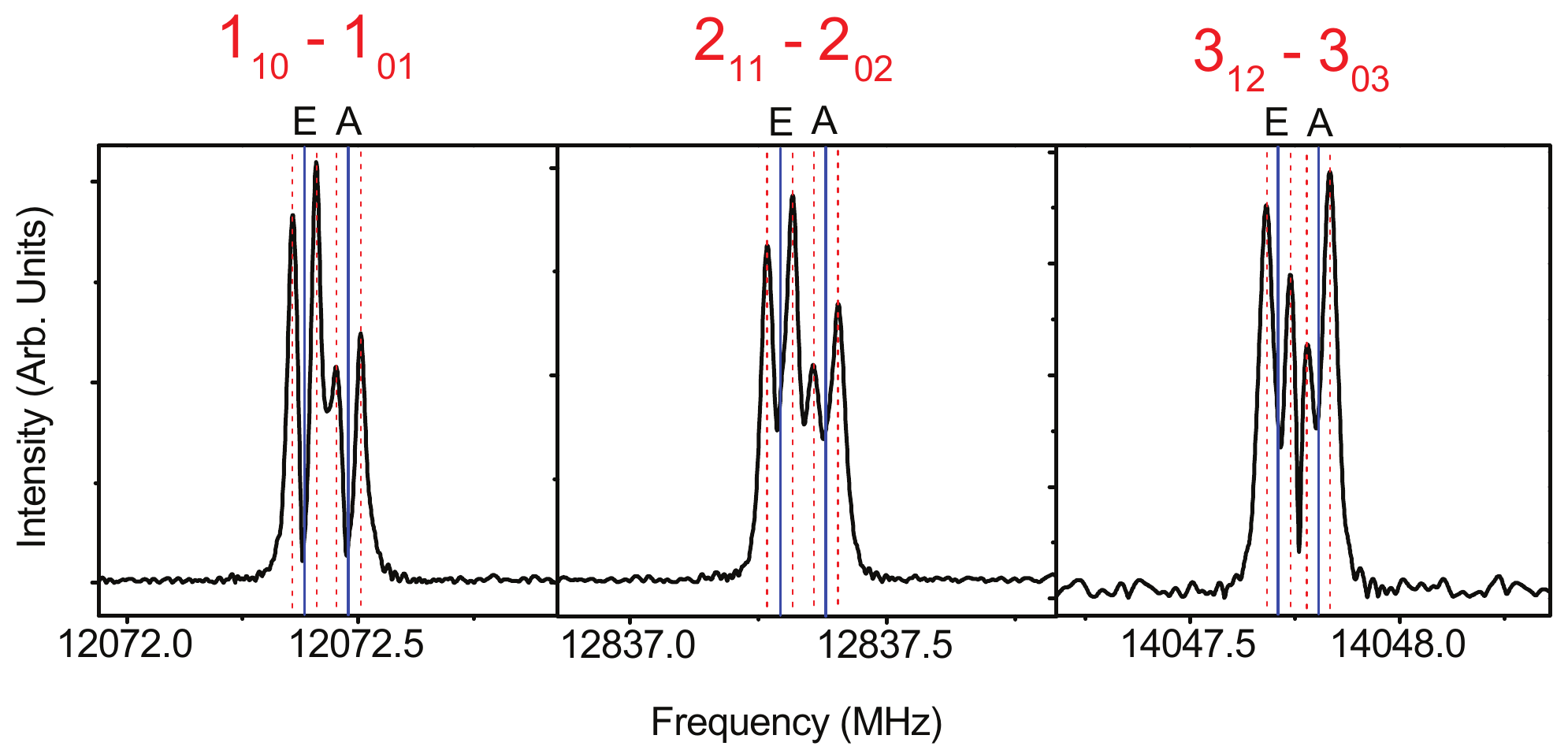}
\caption{\textbf{Laboratory spectra of astronomically-observed propylene oxide transitions.}  Laboratory measurement of the three transitions detected astronomically in this work, showing the characteristic A-E methyl rotor splitting (solid blue lines) and Doppler splitting (dashed red lines). Rotational quantum numbers are given in red.}
\label{propoxzoom}
\end{figure}

The use of the same local oscillator (LO) for both up and downconversion gives double sideband spectra, thus the absolute the frequency of any emission cannot be determined from a single scan. To determine absolute frequencies, a second scan is recorded at a slightly shifted (LO+10 MHz) LO frequency. The deconvolved broadband spectrum is shown in Figure \ref{propox} using a 10 microsecond FID. To resolve the A-E splitting, we followed up with targeted measurements using a 130 microsecond FID (Figure \ref{propoxzoom}). Each peak is a doublet of doublets; the A-E methyl rotor splitting is $\sim$100 kHz, while the Doppler splitting from the jet expansion is $\sim$50 kHz. Fit frequencies are given in Table \ref{freqs}.

\begin{table}
\centering
\caption{\textbf{Measured laboratory frequencies of the transitions used in this work, and associated observed linewidth broadening.}}
\label{freqs}
\begin{tabular}{c c c c }
\hline\hline
J$^{\prime}_{K_aK_c}$ -- J$^{\prime\prime}_{K_aK_c}$	&	Symmetry		& 	Frequency (MHz) 	&	$\Delta V$ (km s$^{-1}$)	\\
\hline
$1_{10} - 1_{01}$								&	E			&	12072.384(20)		&	\multirow{2}{*}{2.4}\\
											&	A			&	12072.479(20)		\\
$2_{11} - 2_{02}$								&	E			&	12837.292(20)		&	\multirow{2}{*}{2.1}\\
											&	A			&	12837.381(20)		\\
$3_{12} - 3_{03}$								&	E			&	14047.715(20)		&	\multirow{2}{*}{1.9}\\
											&	A			&	14047.806(20)		\\	
\hline
\end{tabular}
\end{table}
	
\clearpage

\section*{The Observational Model}

To determine a column density and rotational temperature for propylene oxide, we follow the convention of \cite{Hollis:2004uh}, using Equation \ref{cd_eqn} to calculate the column density $N_T$, given a rotational partition function $Q_r$, upper state energy $E_u$, rotational excitation temperature $T_{ex}$, transition frequency $\nu$, line strength $S\mu^2$, observed intensity $\Delta T_A^*$, linewidth $\Delta V$, telescope efficiency $\eta_B$, and background temperature $T_{bg}$.
\begin{equation}
N_T = \frac{Q_re^{E_u/kT_{ex}}}{\frac{8\pi^3}{3k}\nu S\mu^2}\times \frac{\frac{1}{2}\sqrt{\frac{\pi}{ln(2)}}\frac{\Delta T_A^*\Delta V}{\eta_B}}{1 - \frac{e^{h\nu/kT_{ex}} -1}{e^{h\nu/kT_{bg}} -1}}
\label{cd_eqn}
\end{equation}
We make the assumption that all propylene oxide states can be described by a single excitation temperature equal to $T_{ex}$.  We stress that this assumption does not imply that the molecule is necessarily in thermal equilibrium with the gas; $T_{ex}$ is not assumed to be equal to the ambient gas kinetic temperature, a condition often described as Local Thermodynamic Equilibrium, or LTE. The excitation of molecules in Sgr B2(N), especially those in the cold absorbing layer(s) as is propylene oxide, rarely can be described by the gas kinetic temperature; yet, a single-excitation model often describes such species well \cite{Hollis:2004uh,Loomis:2013fs,Zaleski:2013bc}.

The rotational partition function, $Q_r$, is often calculated according to a high-temperature approximation, given by  Equation \ref{Q} where $\sigma$ is a unitless symmetry parameter, $T_{ex}$ the excitation temperature (K), and $A$, $B$, and $C$, the rotational constants of the molecule (MHz) (c.f. \cite{Gordy:1984uy}), which offers excellent values down to modestly-low temperatures.  
\begin{equation}
Q_r = \left(\frac{5.34 \times 10^6}{\sigma}\right)\left(\frac{T_{ex}^3}{ABC}\right)^{1/2}
\label{Q}
\end{equation}
Molecules seen in absorption toward Sgr B2(N), however, often are characterized by $T_{ex}$ values as low as 6--8 K \cite{Hollis:2004uh,Hollis:2006ih}, in which case direct summation of the energy levels is required, as given in Equation \ref{summation} (c.f. \cite{Gordy:1984uy}).
\begin{equation}
Q_r = \frac{1}{\sigma}\sum_{J = 0}^{J = \infty}\sum_{K = -J}^{K = J}(2J+1)e^{-E_{J,K}/kT_{ex}}
\label{summation}
\end{equation}
For asymmetric molecules like propylene oxide, the symmetry parameter $\sigma = 1$.  In this study, we directly sum the rotational states using Equation \ref{summation} to determine $Q_r$, however, we note that at the temperatures under consideration, the error in Equation \ref{Q} is only 1.3\% at 6 K and drops below 0.1\% by 40 K.

The $S\mu^2$ value is the intrinsic line strength $S_{ij}$ multiplied by the square of the transition dipole moment $\mu$, which in the case of the pure rotational transitions considered here is simply the permanent electric dipole moment $\mu$ along the principal axis of the transition.  The $S_{ij}$ factors are intrinsic quantum mechanical properties determined from the $J$ and $K$ values of the transition; a thorough reference tabulation is available \cite{Townes:1975ve}.  The dipole moments $\mu_a$, $\mu_b$, and $\mu_c$ were taken from \cite{Swalen:1957fn} as 0.95, 1.67, and 0.56 Debye, respectively.

The observed intensities, $\Delta T_A^*$, and linewidths, $\Delta V$, were obtained by fitting a single Gaussian lineshape to the observed transitions (for PRIMOS observations), or a pair of Gaussians (for Parkes observations).  In either case, the observed signals are well-fit. The results are given in Table \ref{g_fits}.  Given the uncertainties, we adopted a uniform linewidth of 13 km s$^{-1}$ when considering the PRIMOS transitions.  We note a trend of decreasing linewidth with increasing frequency, which we attribute to a combination of the decreasing beam size encompassing less inhomogeneous environments, and the decreasing broadening due to internal rotation as J-values increase.
\begin{table}[h!]
\centering
\caption{\textbf{Line parameters resulting from Gaussian Fits to the observed transitions.} Numbers in parenthesis are 1$\sigma$ standard deviations in units of the last significant digit.}
\begin{tabular}{c  c c c c c c c}
\hline\hline
				&	\multicolumn{3}{c}{Component 1}					&\hspace{0.1in} 	&	\multicolumn{3}{c}{Component 2}\\
\cline{2-4}\cline{6-8}
				&	$v_o$		&	$\Delta T_A^*$	&	$\Delta V$		&\hspace{0.1in}		&	$v_o$		&	$\Delta T_A^*$	&	$\Delta V$	 	\\
Transition			&	(km s$^{-1}$)	&	(mK)			&	(km s$^{-1}$)	&				&	(km s$^{-1}$)	&	(mK)			&	(km s$^{-1}$)	\\
\hline
$1_{1,0} - 1_{0,1}$	&	59.3(18)			&	-45(3)	&	19.6(33)		&				&	45.5(23)		&	-13(9)		&	9.9(57)		\\
$2_{1,1} - 2_{0,2}$	&	61.4(7)			&	-7.4(5)	&	15.8(20)		&				&	...			&	...			&	...			\\
$3_{1,2} - 3_{0,3}$	&	63.0(8)			&	-6.8(6)	&	11.6(14)		&				&	...			&	...			&	...			\\
\hline
\end{tabular}
\label{g_fits}
\end{table}

The lines were fit, and the column density and temperature determinations conducted, using the laboratory frequencies ($\nu$) determined without explicitly treating the splitting of the rotational levels due to internal motion because the splitting is not resolved in our observations.  This splitting does, however, contribute a non-negligible amount to the total linewidth of the observed features: 2.4, 2.1, and 1.9 km s$^{-1}$ at 12.1, 12.8, and 14.0 GHz, respectively.  

The telescope beam efficiency ($\eta _b$) was calculated explicitly at each frequency across the PRIMOS band using Equation \ref{eta} (c.f. \cite{Hollis:2007ww}).
\begin{equation}
\eta_b = -15.52 \times 10^{-5}\nu ^2 - 22.59 \times 10^{-4} \nu + 0.98
\label{eta}
\end{equation}
As we also include the observations of \cite{Cunningham:2007kl} in our analysis, which were acquired with the MOPRA telescope, we take an average value of $\eta_b = 0.44$ uniformly across the 3 mm MOPRA observational window \cite{Ladd:2013hn}.

The background temperature ($T_{bg}$) plays a critical role in the accurate modeling of molecular excitation for transitions searched for in the PRIMOS observations, in particular for those molecules which are seen in absorption.  This is readily apparent if Equation \ref{cd_eqn} is re-stated as Equation \ref{cd_abs}, below (c.f. \cite{Hollis:2004uh}).
\begin{equation}
N_T = \frac{Q_r\frac{1}{2}\sqrt{\frac{\pi}{ln(2)}}\frac{\Delta T_A^* \Delta V}{\eta _B}}{\frac{8\pi^3}{3h}(T_{ex}-\frac{T_{bg}}{\eta_B})S\mu^2(e^{-E_l/kT_{ex}}-e^{-E_u/kT_{ex}})}
\label{cd_abs}
\end{equation}
Here, it is clear from the $T_{ex}-\frac{T_{bg}}{\eta_B}$ term that for values of $T_{ex} < T_{bg}$, $T_A^*$ must be negative (in absorption) for $N_T$ to remain positive.  As the observed propylene oxide transitions are in absorption, it is therefor critical to constrain $T_{bg}$ in Sgr B2(N) at the frequencies observed.  In this case, a model must first be adopted to describe the overlap of the GBT beam with the background continuum emission structure at each frequency.

The structure of Sgr B2(N) is complex, with a compact ($\sim$$5^{\prime\prime}$) hot molecular core surrounded by a more extended, colder molecular shell \cite{Hollis:2007ww}.  The  background continuum structure against which molecules in this shell absorb is $\sim$20$^{\prime\prime}$ in diameter (see, e.g., Fig. $3b$ of \cite{Mehringer:1993dd}).  For the purposes of this study, we assume this 20$^{\prime\prime}$ source size for the cold molecular material, and explicitly calculate the spatial overlap of these regions with the GBT beam as its size varies across the frequency coverage.  The geometry-corrected source size is then used to determine a correction factor ($B$) to observed intensities for beam dilution effects, according to Equation \ref{dilution}, where $\theta_s$ and $\theta_b$ are the source and beam sizes, respectively.  This is then applied as required to both peak line intensities ($\Delta T_A^*$) and to the background continuum levels, as described below.

\begin{equation}
B = \frac{\theta_s^2}{\theta_s^2 + \theta_b^2}
\label{dilution}
\end{equation}

The background continuum temperature at 85 points across the PRIMOS frequency coverage was established by \cite{Hollis:2007ww}.  These continuum measurements range from more than 100 K at low frequencies to several Kelvin above the CMB near 45 GHz, although these measurements are not corrected for any assumed source geometry or beam dilution effects.  The authors attribute this to non-thermal continuum in the source.  We adopt their measurements of the continuum for these calculations, but correct them for the overlap of the assumed source size of the continuum-emitting region with the GBT beam in the PRIMOS observations.  We expect this correction to be valid to frequencies as low as $\sim$10 GHz, below which the GBT beam is sufficiently large that it will encompass additional emitting regions.  At that point, the beam dilution correction factor used here will no longer account for the full size of the emitting regions within the beam, and the continuum will be over-estimated in increasing degree. 

The effect of this frequency-dependent continuum on the observed intensities of transitions, both in emission and in absorption, is significant.  While the relative populations of the energy levels for a molecule at a single excitation temperature are still described by a Boltzmann distribution, the observed relative intensities of the transitions between those levels are not.  This is most straightforward to understand looking at absorption transitions.  The standard analysis of relative intensities from a molecule whose population is described by a Boltzmann distribution at a single excitation temperature provides predicted absorption depths (i.e. percent absorption).  The measured (observed; $\Delta T_A^*$) value is not, however, directly comparable, as the background against which a 10\% absorptive transition at 45 GHz is absorbing is not the same as the background against which a 1\% absorptive transition at 10 GHz is absorbing.  Thus, the predicted absolute absorption spectrum for propylene oxide in the PRIMOS observations will present different relative intensities between transitions, due to this non-constant background, than a simple percentage absorption spectrum.

\section*{Model Results}

\begin{figure}[h!]
\centering
\includegraphics[width=\textwidth]{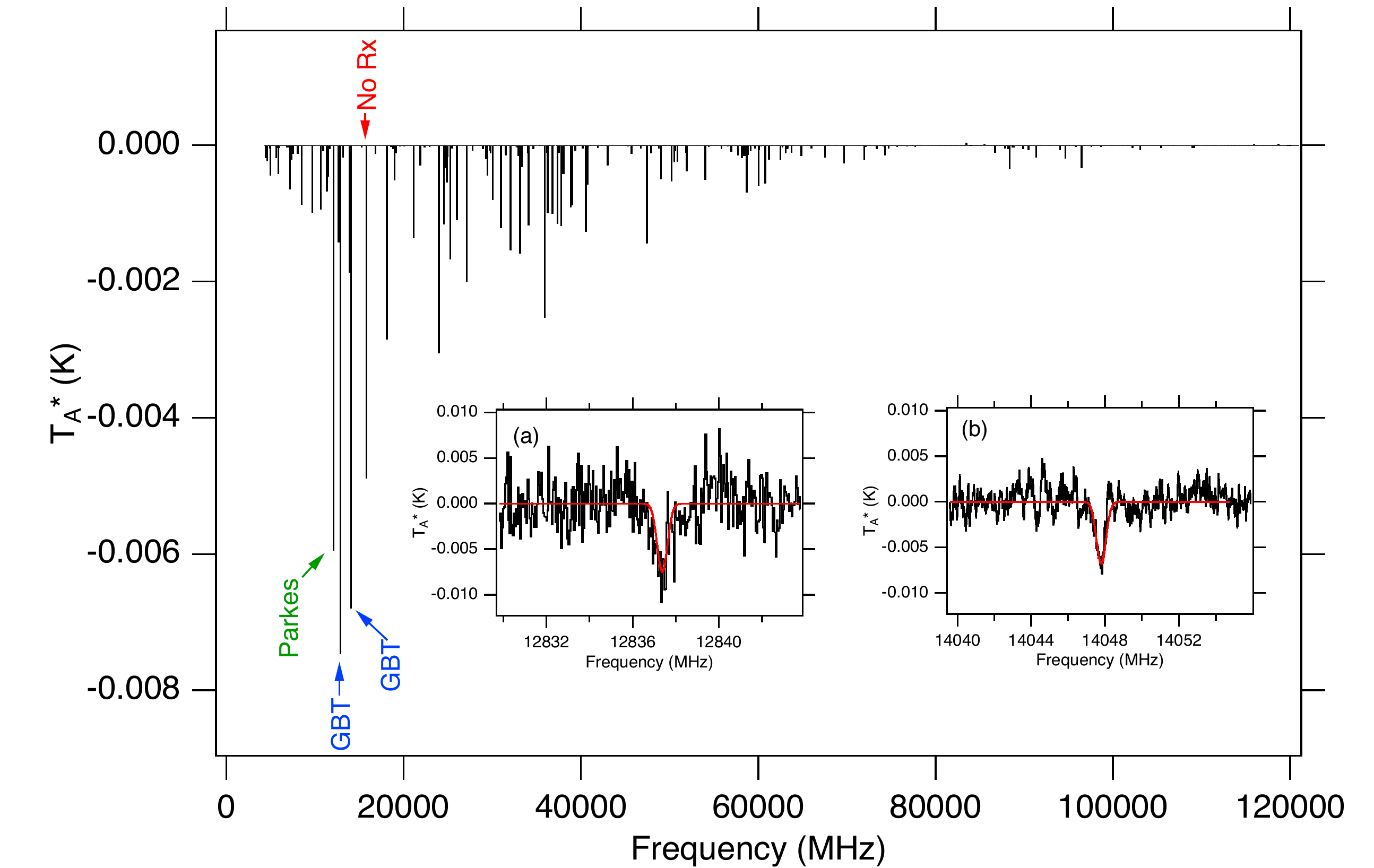}
\caption{\textbf{Model spectrum of propylene oxide toward Sgr B2(N) at the best-fit column density and temperature, and comparison with observations.} The spectrum of propylene oxide at $N_T = 1 \times 10^{13}$ cm$^{-2}$ and $T_{ex} = 5.2$ K, corrected for background continuum, telescope-specific parameters, and beam dilution effects is shown in the main figure. The observed transitions in PRIMOS are shown in insets (a) and (b), with the model spectra overlaid in red.  The next strongest transition has no corresponding receiver (Rx) at the GBT or Parkes, and is marked as such.}
\label{5ksim}
\end{figure}

Using the model described above, we have performed a least-squares analysis of our data, taking into account the two observed transitions in PRIMOS, all transitions covered by PRIMOS but for which no features are seen, as well as those transitions in the coverage of the survey conducted by \cite{Cunningham:2007kl}.  Due to the large difference beam size, telescope-specific parameters, receiver efficiencies, and calibration uncertainties, the Parkes transition was not included in the analysis quantitatively.  The results of the analysis provide a set of models for column density and temperature which are consistent with our observations, meaning that given the frequency coverage and sensitivity of the PRIMOS observations and those of \cite{Cunningham:2007kl}, only the transitions we have detected are expected to be present above the noise level of the observations.  The best-fit model for the datasets, including both detected and non-detected lines in the PRIMOS observations and those of \cite{Cunningham:2007kl}, is for a column density of $1 \times 10^{13}$ cm$^{-2}$ and an excitation temperature of $T_{ex} = 5.2$ K.  A simulated spectrum under these conditions is shown in Figure \ref{5ksim}, while insets (a) and (b) show simulations of the model overlaid on the observational spectra.

The uncertainties for the Gaussian-fitted parameters given in Table \ref{g_fits} are not the dominant source of uncertainty in our analysis.  To quantitatively determine the uncertainty in the column density and excitation temperature, we have followed the approach of \cite{Crockett:2014er}, which takes into account uncertainty in the baseline offset, the local rms noise level, the absolute flux calibration, the pointing error, error in the beam filling factor due to uncertainty in the source size, and, in our case, the uncertainty in the determined continuum level.  We make conservative estimates of these quantities in our analysis, and assume a 20\% uncertainty in the absolute flux calibration, 2$^{\prime\prime}$ uncertainty in the pointing, a 20\% uncertainty in the source size, and a 20\% uncertainty in the continuum level.  Given the sparsity of spectral features, we assume no contribution to the error from the removal of the baseline.

Given these uncertainties, and combined with the two fitted transitions from PRIMOS, there are a range of models which fit the observed transitions, and non-detected transitions, nearly as well spanning excitation temperatures as high as $\sim$35 K.  Nevertheless, a  column density of $1 \times 10^{13}$ cm$^{-2}$ and an excitation temperature of $T_{ex} = 5.2$ K are indeed the best-fit to our observations, although there are a wide range of models which remain consistent with both the detected and non-detected lines from our observations and those with the MOPRA telescope\cite{Cunningham:2007kl}.

\part*{}
\section*{Astronomical Statistical Analysis}

The low line density in the cm-wave region of the spectrum allows for the definitive identification of new molecules from far fewer transitions than are required in the (sub-)mm region.  This is especially true in the case of Sgr B2(N), where at frequencies $>$100 GHz, broad line-widths, high degrees of molecular-complexity, and large column densities contribute to an essentially baseline-free, completely line-confused spectrum.  To illustrate the relative line densities of these two frequency regimes, we have compared our PRIMOS observations around 13 GHz to a selection of spectra toward Sgr B2(N) from the Barry E. Turner Legacy survey taken with the NRAO 12 m telescope on Kitt Peak (accessible at http://www.cv.nrao.edu/$\sim$aremijan/PRIMOS).  The frequency windows were chosen to provide equivalent coverage in velocity space ($\sim$6000 km s$^{-1}$) in both scans.

In the Turner Survey, which has an RMS of $\sim$14 mK, there are $\sim$315 lines at $3\sigma$ or above, with a resulting line density of $\sim$1 line per 10 MHz ($\sim$1 line per 20 km s$^{-1}$). Herschel HIFI and ALMA data have even lower noise floors and are essentially line-confusion limited at all frequencies. Typical linewidths in this source are 8 -- 25 km s$^{-1}$ FWHM, making significant line blending inevitable \cite{Hollis:2006ih,Hollis:2004uh,Loomis:2013fs,Zaleski:2013bc}.  To claim a detection of a line in the presence of blending, however, the components must be separated by at least their FWHM\cite{Snyder:2005tr}, and the coincidence of lines separated by this criterion is greatly reduced in such crowded spectra.  Thus detections at (sub-)mm wavelengths regularly require dozens or hundreds of lines to be secure, and fitting techniques must explicitly account for line blending.

In the PRIMOS survey, there are $\sim$26 lines above $3\sigma$ in the same $\sim$6000 km s$^{-1}$ window, resulting in a similar line density of $\sim$1 line per 10 MHz in frequency space.  In velocity space, however, the line density drops to $\sim$1 line per 230 km s$^{-1}$.  With an average line separation $>10$ times the linewidth, there is almost no chance for coincidental overlap.  Thus as long as the assignments of laboratory spectra are robust, molecular detections at cm wavelengths require only a few (2 -- 5) lines to be secure.

The lack of line confusion in the PRIMOS survey allows for further quantitative constraints on the likelihood of a spurious detection (\emph{13}). Previous observations gave a propylene oxide column density upper limit of $\leq$6.7 $\times$ 10 $^{14}$ cm$^{-2}$, so we may conservatively only consider features between $-$100 mK and $+$100 mK. If we further restrict ourselves to only absorption features, as would be expected for a rotationally cold molecule toward a strong background source, the number of candidate features is $\sim$ 0.28 per 10 MHz. For this line density, the probability of finding a feature coincident with a propylene oxide transition rest frequency, defined conservatively as being within twice the largest source FWHM, 50 km s$^{-1}$ (2.2 MHz), is $\sim$ 0.06. The likelihood of three such coincidences is then $\sim$2$\times$10$^{-4}$. 

The high resolution of our observations provides even further constraint, as all three observed transition peaks fall within 3 resolution elements (73.2 kHz or 1.7 km s$^{-1}$) of the measured laboratory transition frequency. For the given line density, the likelihood of a single coincident transition falling no more than 73.2 kHz from the rest frequency is $\sim$ 0.004, and the likelihood of three transitions all occurring within this window is $\sim$ 6 $\times$ 10$^{-8}$. 

Finally, although a wide range of linewidths are observed toward Sgr B2(N), the two transitions measured with the GBT show nearly identical linewidths, providing further evidence that the transitions are truly related, and are not spurious. As previously noted, the Parkes data samples a substantially different total population and its linewidth cannot be directly compared. 

\section*{Measurement of Circular Dichroism}
The most straightforward method available for astronomical detection of e.e. is circular dichroism. Circular dichroism (CD), or preferential absorption of left- or right-handed circularly polarized light (CPL) manifests itself as a change in the difference in the electric field of left versus right-handed CPL, commonly referred to as the Stokes V parameter,  that follows the absorption profile of the observed (in this case, propylene oxide) features. In principle, detection of e.e. requires two CD measurements. First, is a quantitative laboratory measurement of CD to relate e.e. to an observed CD. Second, and most critically, is a polarization-sensitive astronomical observation to detect CD. 

For laboratory measurements, this requires experimental measurement and quantification of CD for each of the transitions in question. In the present case this would require laboratory absorption measurement of CD for the cm-wave propylene oxide transitions at 12.1,12.8, and 14.0 GHz. To date, CD has been shown in a laboratory setting from the ultraviolet to the mid-infrared.  Extension to the microwave, or radio region, has been studied theoretically\cite{Salzman:1998jd}. These studies conclude that such measurements should be feasible, although they have not yet been experimentally demonstrated.  Such measurements would then enable quantitative analysis of polarization-sensitive astronomical observations of CD discussed below.

Modern radio telescopes, especially unblocked off-axis designs such as the GBT, are capable of highly-accurate, polarization-sensitive observations across wide frequency windows, simultaneously determining the polarization state at each observed frequency (e.g.\cite{Mason:2009dr}). Briefly, this is achieved by using receivers that simultaneously detect the electric field of two orthogonal polarizations. The in-phase (0$^{\circ}$) and quadrature-phase (90$^{\circ}$) components of the electric field for each polarization are then separated and digitized. Appropriate combinations of these four signals are then used to fully determine the polarization state at each frequency within the observed pass band, producing a Stokes vector (discussed below) at each frequency. Acquiring accurate, phase-calibrated, data across large pass bands requires careful calibration, but is regularly achieved by properly designed radio receivers \cite{Robishaw:2009hu}. Such observations are vital both for detection of circular dichroism and to distinguish other effects that may change the polarization state of the detected light, obfuscating potential CD signals. 

In astronomical observations, there are two potential sources of confusion for the detection of circular dichroism: non-resonant effects, i.e. those that do not require light at a frequency corresponding to transitions between states of a molecule, and resonant effects that do. First are macroscopic or non-resonant effects, e.g. Faraday rotation or dust scattering. For such effects, small changes in frequency produce effectively no difference in the effect on the polarization\cite{2011piim.book.....D}. In the present observations, the linewidth is $\sim$ 700 kHz, at frequencies of several GHz, making the difference between on and off resonance is extremely small ($\sim$10$^{-5}$). Therefore any non-resonant effect that alters the polarization at the frequency of the absorption will have the same effect in non-absorption channels and this effect can be corrected by comparison of the polarization on and off resonance. 

The second possible source of confusion is resonant effects. Because resonant effects are specific not only to a molecule, but to a transition between states of a particular molecule, they occur only in the narrow region where a molecule absorbs or emits. Effects that produce changes in linear polarization, e.g. the Goldreich-Kylafis effect\cite{Goldreich:1981ab}, are readily distinguished from circular dichroism, as they produce changes in linear, rather than circular polarization, which is distinct from CD. The plausible effect other than circular dichroism to produce changes in circularly polarized light is the Zeeman effect. In this scenario an applied magnetic field lifts the degeneracy of the \textit{m} sublevels, and positive  \textit{m} sublevels interact preferentially with one handed of circularly polarized light, while negative \textit{m} sublevels interact with the opposite handedness. This too is easily distinguished from circular dichroism. In order to produce non-canceling circular polarization, the Zeeman splitting must be spectrally-resolved.  Spectrally-resolved Zeeman splitting produces distinct peaks with circular polarizations of opposite sign. Conversely, circular dichroism produces a single peak of a single sign\cite{2008ApJ...680..981R,Salzman:1998jd}. 

Furthermore, observing a wide bandwidth allows for additional tests to prevent false positives and constrain polarization effects. By observing one of several achiral species originating from the same region as propylene oxide, e.g. propanal or acetone, it can be confirmed that there are no resonant molecular effects or observational artifacts causing changes in the polarization. Higher order effects such as the Cotton-Mouton effect, quadratic field induced optical activity, and the magnetochiral effect scale nonlinearly with static magnetic field and given the modest measured field strengths toward Sgr B2(N), are not considered\cite{Wagniere:1999ab,Crutcher:1996ab}.

For astronomical observations, it is then critical to demonstrate that CD produces observable polarization effects that are distinct from other potential effects, and is therefore an astronomically observable effect. The polarization state of light can be completely specified by four parameters, commonly chosen to be the four Stokes parameters, I, Q, U, and V, where I is the total intensity, Q is linear polarization, U is 45$^\circ$ linear polarization, and V is circular polarization. This is often written as a vector \textbf{S} = [I,U,Q,V]. Using these vectors, any polarization-altering effect may evaluated using the so-called Mueller matrices\cite{Bass:2009:HOT:1594759}. For on and off-resonance observations, the polarization vectors will be of the form \textbf{V$^\prime$} = \textbf{M$_{los}$}$\cdot$\textbf{V}, where \textbf{V} is some initial Stokes vector, \textbf{M$_{los}$} is a Mueller matrix describing subsequent polarization changes along the line of sight, and \textbf{V$^\prime$} is the Stokes vector measured by the telescope. In the absence of CD, the Stokes vector is exactly as above, while on resonance and additional term describing CD must be included. The resulting vector is then given by \textbf{V$^{\prime\prime}$} = \textbf{M$_{los}$}$\cdot$\textbf{M$_{CD}$}$\cdot$\textbf{V}. The difference between on and off resonance observations, \textbf{V$^{\prime\prime}$}-\textbf{V$^{\prime}$}, is given by \textbf{M$_{los}$}$\cdot$\textbf{V}-\textbf{M$_{los}$}$\cdot$\textbf{M$_{CD}$}$\cdot$\textbf{V} =  \textbf{M$_{los}$}(V-\textbf{M$_{CD}$}$\cdot$\textbf{V}). If these two vectors are identical the difference is zero, and they indistinguishable, as is the effect of circular dichroism. This expression is zero for two conditions, if the kernel of \textbf{M$_{los}$} is not null, or if \textbf{V} is an eigenvector of \textbf{M$_{CD}$}, with eigenvalue 1. The first condition can only be met if the line-of-sight transmittance is zero, i.e. if there is complete absorption along the line of sight. For the second condition \textbf{M$_{CD}$} has eigenvectors [1,0,0,1], [0,1,0,0], [0,0,1,0], [1,0,0,-1]\cite{Bass:2009:HOT:1594759}. Of these four vectors, only two can be physically meaningful, corresponding to the pure circularly polarized states. \textbf{M$_{CD}$} has eigenvalue 1 only for complete transmittance. The result is that the only valid eigenvector for \textbf{M$_{CD}$} with eigenvalue 1 is completely transmitted pure circular polarization. Thus the only two cases in which the on and off resonance Stokes vectors are indistinguishable are total absorption or the absence of any CD, meaning that CD must produce a distinct Stokes vector. 

From this analysis it is clear that observations at a frequency where CD occurs and where it does not must produce distinct results, and careful measurements may thus identify CD in observations. In the case of the present observations however, a meaningful measurement of the Stokes vector is not possible. This is principally due to a lack of polarization calibration. Even extremely well-designed receivers such as those at the GBT, may introduce significant polarization artifacts through a variety of effects, including polarization side lobes, beam squint, and non-orthogonal response \cite{Robishaw:2009hu}. These effects have dependence on frequency, position angle throughout observations, altitude and azimuth coordinates of the dish, and even cable length, and therefore necessitate great care to accurately determine Stokes vectors and prevent detection of spurious polarization effects.

\section*{Propanal and Acetone Observations in PRIMOS}

We have examined the PRIMOS dataset for transitions of acetone and propanal.  We find 18 clear, unblended transitions of acetone (Table \ref{acetone}), and 11 similarly distinct transitions of propanal (Table \ref{propanal}).  All transitions were observed in absorption, and parameters derived by single-Gaussian fits to the lines were determined.  We use the method described above to fit a column density and temperature to these species using these observed transitions, but note that we make no effort to refine the physical models from those used for propylene oxide.  A static linewidth of 12 km s$^{-1}$ was used for acetone, and a linewidth of 9 km s$^{-1}$ for propanal.  The source size was taken as 20$^{\prime\prime}$ in both cases.

\begin{figure}[h!]
\centering
\includegraphics[width=\textwidth]{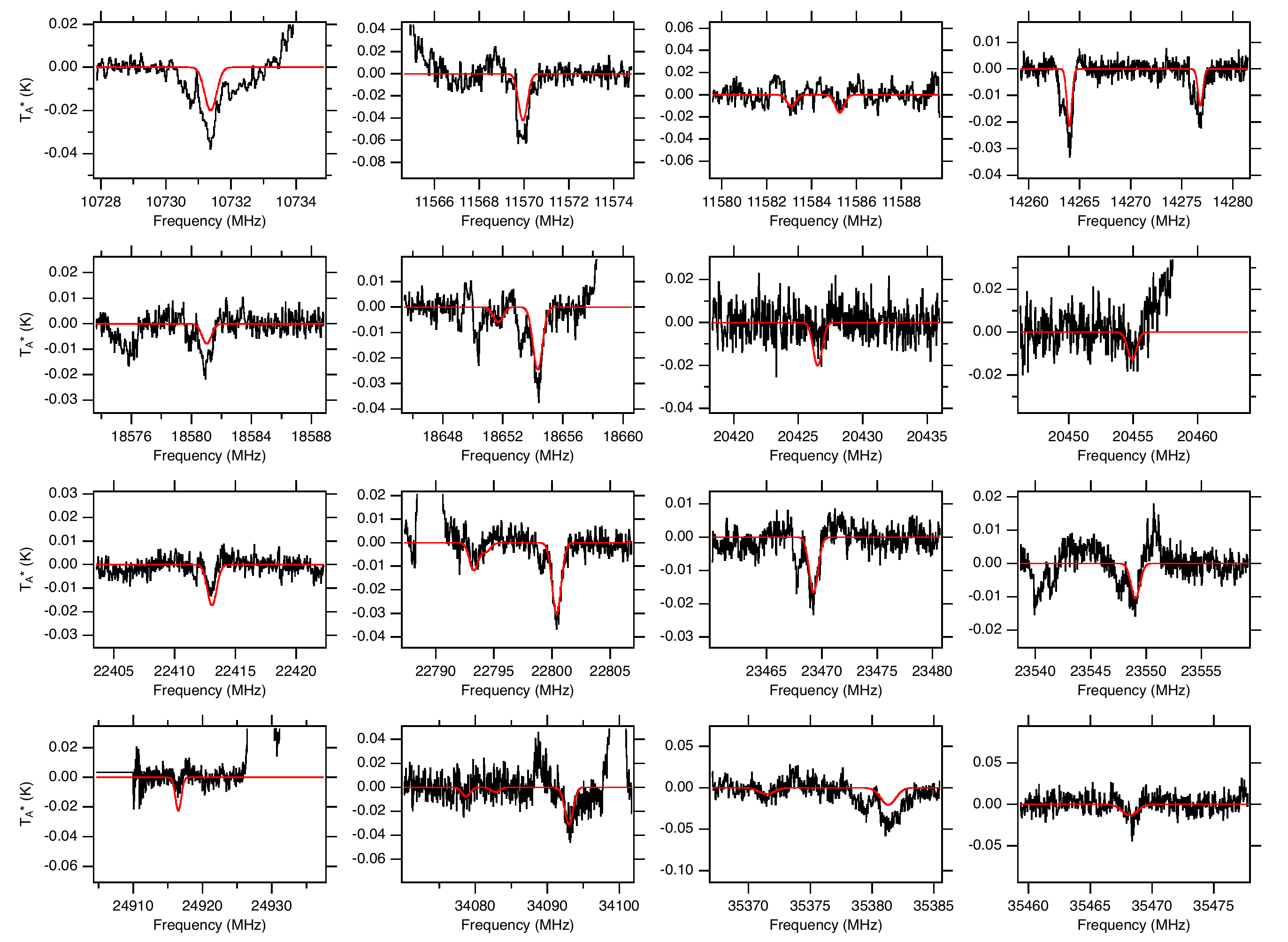}
\caption{\textbf{Model spectrum of acetone toward Sgr B2(N) at the best-fit column density and temperature, and comparison with observations.} The spectrum of acetone at $N_T = 2.1 \times 10^{14}$ cm$^{-2}$ and $T_{ex} = 6.2$ K, corrected for background continuum, telescope-specific parameters, and beam dilution effects is shown in the main figure. The observed transitions in PRIMOS are shown in black, with the model spectra overlaid in red.}
\label{acetonefig}
\end{figure}

\begin{figure}[h!]
\centering
\includegraphics[width=\textwidth]{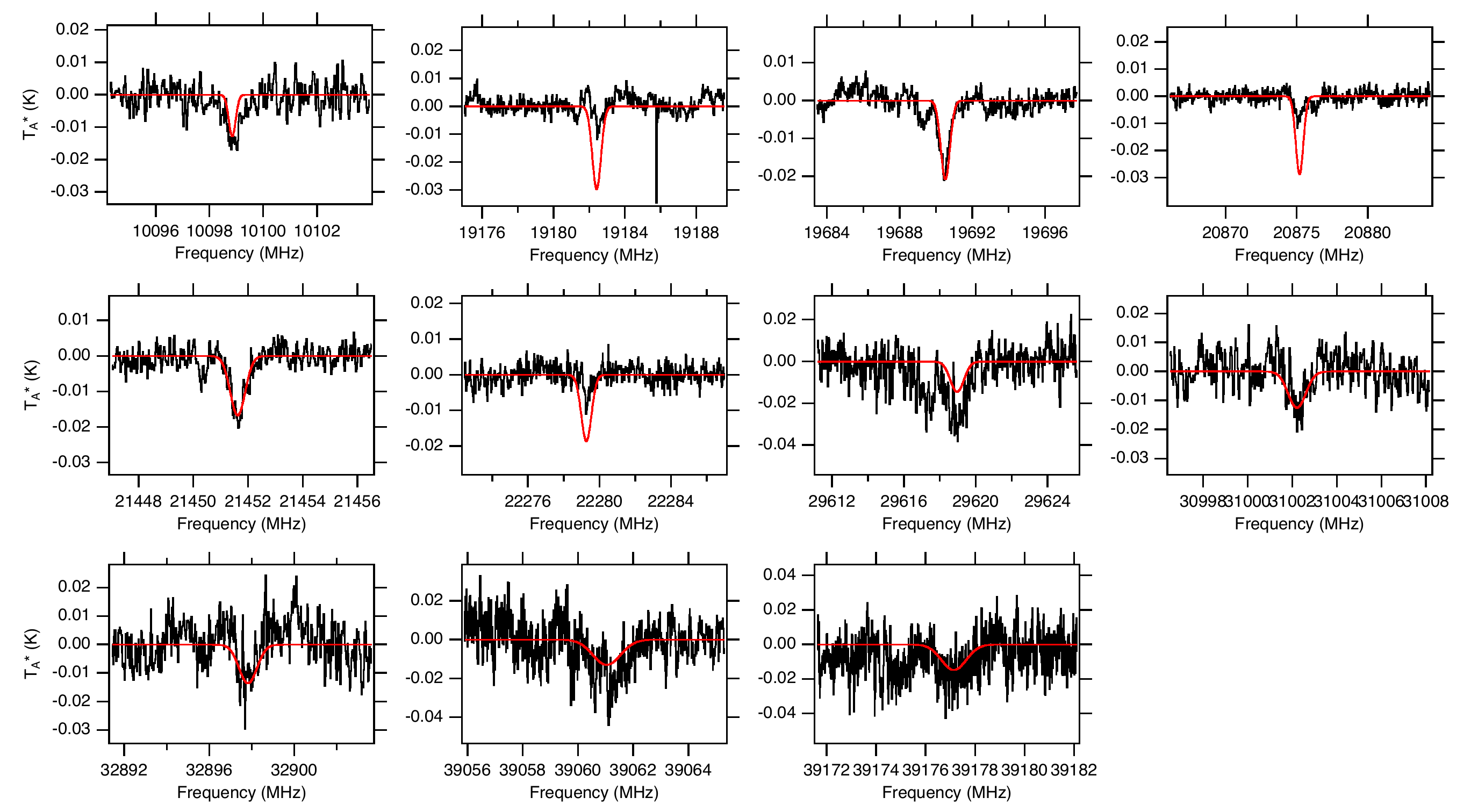}
\caption{\textbf{Model spectrum of propanal toward Sgr B2(N) at the best-fit column density and temperature, and comparison with observations.} The spectrum of propanal at $N_T = 6 \times 10^{13}$ cm$^{-2}$ and $T_{ex} = 6.2$ K, corrected for background continuum, telescope-specific parameters, and beam dilution effects is shown in the main figure. The observed transitions in PRIMOS are shown in black, with the model spectra overlaid in red.}
\label{propanalfig}
\end{figure}

The models derived here reproduce the observed intensities to within a factor of $\sim$2 (Figures \ref{acetonefig} and \ref{propanalfig}), and the uncertainties in these parameters should be taken to be of this order.  Further, for acetone in particular, a warm component is known to be present within the PRIMOS beam, but was not explicitly treated in this analysis.  Thus, while the derived values are useful for qualitative comparisons with each other and propylene oxide, a more rigorous treatment than the single population, single excitation-temperature model used here, although beyond the scope of this work, would refine the results. 

\begin{table}
\caption{\textbf{Observed acetone transitions in PRIMOS.}}
\label{acetone}
\centering
\begin{tabular}{c c c c}
\hline\hline
$J^{\prime}_{K_a,K_c} - J^{\prime\prime}_{K_a,K_c}$	&	Frequency (MHz)	&	$\Delta T_A^*$ (mK)		&	$\Delta V$ (km s$^{-1}$)	\\
\hline
$3_{2,1} - 3_{1,2}$ AE							&	10731.360(9)		&	-33.0(9)				&	16.5(6)				\\
$4_{3,1} - 4_{2,2}$ EE							&	11569.943(4)		&	-60(3)				&	14.9(9)				\\
$4_{3,1} - 4_{2,2}$ EA							&	11583.080(5)		&	-13(3)				&	10(3)					\\
$4_{3,1} - 4_{2,2}$ AA							&	11585.245(3)		&	-11(2)				&	10(3)					\\
$5_{4,1} - 5_{3,2}$ EE							&	14263.975(5)		&	-28.7(8)				&	15.8(6)				\\
$5_{4,1} - 5_{3,2}$ AA							&	14276.816(6)		&	-18.6(7)				&	14.8(9)				\\
$4_{2,2} - 4_{1,3}$ EE							&	18654.309(4)		&	-29.7(8)				&	12.7(4)				\\
$3_{1,2} - 3_{0,3}$ AA							&	20454.895(4)		&	-9(1)					&	12(2)					\\
$3_{2,2} - 3_{1,3}$ EE							&	22413.051(3)		&	-12.2(5)				&	8.8(4)				\\
$2_{0,2} - 1_{1,1}$ AA							&	22793.262(3)		&	-7.5(7)				&	12(2)					\\
$2_{0,2} - 1_{1,1}$ EE							&	22800.382(2)		&	-32.2(7)				&	11.8(3)				\\
$4_{3,2} - 4_{2,3}$ EE							&	23469.238(4)		&	-18.4(7)				&	12.2(5)				\\
$4_{3,2} - 4_{2,3}$ AA							&	23549.025(4)		&	10.7(9)				&	10.7(9)				\\
$2_{1,2} - 1_{0,1}$ AA							&	24916.487(3)		&	-7.3(9)				&	9(1)					\\
$3_{1,3} - 2_{0,2}$ EE							&	34092.973(3)		&	-23(2)				&	11.6(9)				\\
$2_{2,1} - 1_{1,0}$ AE							&	35371.432(5)		&	-13(2)				&	8(1)					\\
$2_{2,1} - 1_{1,0}$ EE							&	35381.289(4)		&	-43(1)				&	17.8(7)				\\
$2_{2,1} - 1_{1,0}$ AA							&	35468.174(5)		&	-22(2)				&	6.8(9)				\\
\hline
\multicolumn{4}{l}{Numbers in parentheses are 1$\sigma$ uncertainties in units of the last} \\
\multicolumn{4}{l}{significant digit. A and E labels designate torsional sub-levels.}
\end{tabular}
\end{table}

\begin{table}
\caption{\textbf{Observed propanal transitions in PRIMOS.}}
\label{propanal}
\centering
\begin{tabular}{c c c c}
\hline\hline
$J^{\prime}_{K_a,K_c} - J^{\prime\prime}_{K_a,K_c}$	&	Frequency (MHz)	&	$\Delta T_A^*$ (mK)		&	$\Delta V$ (km s$^{-1}$)	\\
\hline
$2_{0,2} - 1_{1,1}$								&	10098.84(5)		&	-14.3(8)				&	21(1)					\\
$4_{1,3} - 4_{0,4}$								&	19182.410(4)		&	-8.3(8)				&	5.8(8)				\\
$2_{1,2} - 1_{1,1}$								&	19690.51(5)		&	-18.2(4)				&	10.2(3)				\\
$2_{0,2} - 1_{0,1}$								&	20875.19(5)		&	-8.6(5)				&	8.7(5)				\\
$3_{0,3} - 2_{1,2}$								&	21451.62(5)		&	-17.0(7)				&	9.0(4)				\\
$2_{1,1} - 1_{1,0}$								&	22279.26(5)		&	-8.6(9)				&	3.8(4)				\\
$4_{2,2} - 4_{1,3}$								&	29618.95(5)		&	-28(1)				&	13.8(9)				\\
$3_{2,1} - 3_{1,2}$								&	31002.20(5)		&	-13(2)				&	6.8(9)				\\
$4_{0,4} - 3_{1,3}$								&	32897.82(5)		&	-13(2)				&	6(1)					\\
$3_{1,3} - 2_{0,2}$								&	39061.03(5)		&	-21(1)				&	10.6(8)				\\
$4_{1,4} - 3_{1,3}$								&	39177.132(5)		&	-20(1)				&	12(2)					\\
\hline
\multicolumn{4}{l}{Numbers in parentheses are 1$\sigma$ uncertainties in units of the last} \\
\multicolumn{4}{l}{significant digit.}
\end{tabular}
\end{table}

\end{document}